\documentstyle[preprint,aps,psfig]{revtex}
\begin{document}

%\input psfig

%\twocolumn[

%\hsize\textwidth\columnwidth\hsize\csname @twocolumnfalse\endcsname

\draft

\title{A Barrier Penetration Model for DNA Double Strand Separation}

\author{Ajit Kumar Mohanty}
\address{Nuclear Physics Division,\\
Bhabha Atomic Research Centre, Mumbai-400085, India}

\maketitle

\begin{abstract}
A barrier penetration model has been proposed to explain the temperature driven
spontaneous melting of the DNA oligomers into two separate single
strands whereas the fraction of partially melted
intermediate states can be understood on the basis of bound state solution of the
effective potential obtained using
Peyrard-Bishop formalism. The excellent agreement of the predictions
with the recent experimental measurements
(Euro. Phys. Lett. 62, 452, 2003, Phy. Rev. Lett. 91, 148101, 2003)
provides strong justification for the proposed model. A simple relation
has been obtained
that correlates the strand stiffness with  the temperature
at which the spontaneous melting probability becomes half. This is an important
outcome as the stiffness parameter of the double stranded conformation can be directly
estimated by knowing the mid point temperature which can be measured experimentally.
Further, it is shown that the derivative of this probability with respect to temperature
is quite sensitive to the parameters of the entropic barrier and can be used
as  the experimental probes to study the stacking interaction in detail.

\end{abstract}

%]

%\narrowtext
\section {Introduction}
The DNA molecule is a double-stranded biopolymer with two complementary
sugar-phosphate chains (backbones) twisted around each other forming a right handed
helix \cite{saenger,watson}. Each chain
is a linear polynucleotide consisting of four kinds of bases: Adenine, Guanine
(purines) and Cytosine, Thymine (pyrimidines). The two chains are joined together
by hydrogen bonds between pairs of nucleotides, A-T  with two hydrogen
bonds and G-C with three hydrogen bonds. At sufficiently high
temperature, the double stranded DNA (dsDNA)
helix melts and the molecules separate into two single
strands (ssDNA). The experimental procedure consists in preparing a set of dsDNA
with the same length and sequences, which are immersed in an aqueous solution with
given physiological conditions. Then the fraction of broken base pairs as a function
of temperature, referred to as the melting curve, is measured by the $UV$ light
absorption, typically at about $260~nm$.
Long DNA molecules give rise to steps in the melting curves corresponding to
different regions melting at different temperatures whereas synthetic oligonucleotides
do not show any multistep behavior (see \cite{wartell} for a review).
This denaturation process, often interpreted
as a reminiscent of a discontinuous first order phase transition, has attracted a
great deal of interest to study the physics of highly nonlinear biomolecular system
and also it has led to several micromanipulation experiments
\cite{poland70,peyrard89,dauxois93,dauxois95,zhang97,cule97,bockelmann97,campa98,nyeo99,cocco99,lubensky00,causo00,kafri00,theodorakopoulos00,singh01,cocco02,theodorakopoulos02,lubensky02,dauxois02,carlon02,danilowicz03,antonio03,yin03,richard03}.
 From biophysics point of
view, the strand separation is a key aspect of DNA transcription and
replication \cite{alberts}. During transcription, a transient bubble of ssDNA is
formed to allow the enzyme that makes a RNA copy of the DNA sequence to access
the DNA bases. During replication that occurs during cell division for example, the
dsDNA separates into two, each strand then serving as a template for the synthesis
of a new dsDNA. On the otherhand, accurate prediction of DNAs thermal denaturation
is an important parameter for many biological techniques like polymerases chain
reaction (PCR) where dsDNA are routinely converted to separated ssDNAs by melting
at elevated temperature.  Therefore, the study of DNA denaturation which is
an important biological phenomena has become a subject of interdisciplinary
interest.

The early theoretical models like Poland-Scheraga (PS) type \cite{poland70,poland66,fisher66} considers
the DNA molecule as being composed of an alternating sequence of bound and
denaturated (loop or bubble) states. Typically, a bound state is energetically
favored over an unbound one while a denaturated segment is entropically favored
over a bound one.
The DNA denaturation can be viewed as a competition between the entropy
of the denaturated loops and the bound state energy of the sequence.
In PS type models, the segments that compose the chain are assumed
to be noninteracting with each other. This assumption considerably simplifies the
calculation of free energy. In the past, the entropy of the denaturated loops has
been evaluated by modeling them either as ideal random walks \cite{poland66} or as
self avoiding walks \cite{fisher66}. It has been found that within this approach
the denaturation transition of DNA is continuous both in two and three dimensions,
and it becomes first order only for $d \ge 4$. A denaturated segment in between two
bounded structures may introduce flexibility , allowing neighboring segments to
interact with the bubble and with each other. Inclusion of this excluded volume
interaction effect in PS model drives the transition
first order even for $d=2$ ( and
above) \cite{causo00,kafri00,carlon02,richard03}.
In a different approach, Peyrard and Bishop (PB) \cite{peyrard89} proposed a microscopic
model to describe the dynamics of the DNA denaturation by introducing an interaction
potential which depends on the transverse stretching of the hydrogen bonds between the
complementary base pairs. An improved version of this PB model with more realistic
treatment of the backbone stiffness \cite{dauxois93,dauxois95} yields what appears
to be a first order phase transition. Despite the simple formalism, the PB model
has successfully reproduced the essential features of thermally induced denaturation
of long DNA chains. It has  been used to estimate the melting curves of very
short heterogeneous DNA segments, in excellent quantitative agreement with experimental
data \cite{campa98}. It also provides the characteristic multistep melting behavior
observed in heterogeneous DNA sequence \cite{cule97}. Recently, this model has been used
to investigate charge transport properties in DNA chain \cite{komineas02} and the role
of nonlinearity and entropy provided by the local base pairs on the interaction of the
DNA transcription \cite{kalosakas03}.

As mentioned before, the UV absorption of the spectroscopic measurements are interpreted
as the yield for the average fraction of the open base pairs.
The denaturation process leads to a sharp transition from dsDNA to ssDNA
when the sequence is long. For short oligomers, this transition is
not sharp enough. Due to finite size effect, two fractions,
fully dissociated molecules ($p$) due to
opening of all the base pairs spontaneously
and partially melted intermediate states ($\sigma$), may co-exist over a
wide range of temperature below the melting point.
From the spectroscopic
measurements,  it is not
possible to estimate these two  fractions separately.
Recently, a new method has been proposed \cite{montrichok03} to
quantify the presence of intermediate states and fully open molecules by combining
the UV spectroscopy with the quenching technique. The sequences which are
partially self-complementary are prepared in the duplex form by hybridizing with the
exact reverse complement. On heating, the duplex either melts spontaneously into two
separate single strands or remains as connected but with many broken base pairs.
Molecules which are completely open form hairpins on quenching while the molecules
which are partially open close again as duplexes. The relative amount of hairpins
give the fraction $p$ of completely open molecules whereas the UV spectroscopic measurements
yield the fraction of  total open base pairs $f$. The difference $\sigma= f-p$ represents
the fraction of partially melted intermediate states.
This method has yielded new data on the
fraction of $f$, $p$ and
$\sigma$ and also the relative average bubble size $<l>$ as a function of
temperatures in the melting transition of several DNA oligomers \cite{montrichok03,zeng03}.
In this work, we try to understand this new and exciting experimental data both
qualitatively and quantitatively within the frame work of PB formalism which has been
quite successful elsewhere. We avoid the rigorous solution of the transfer
integral (TI)
equation to evaluate the configurational partition function. Instead, we
concentrate on yet simple but
well approximated TI integral which yields a Schrodinger like second order differential
equation. For simplicity, we use Morse potential \cite{morse29} with inclusion  of
anharmonic stacking interaction \cite{dauxois93}.
We propose here that for the spontaneous melting of the dsDNA leading to a complete
strand separation , the molecules are required
to overcome an effective barrier through thermally assisted
tunneling whereas the fraction of partially
melted intermediate states can be explained on the basis of the bound state solution
of the effective potential in the temperature range for which the bound state exists.
The excellent agreement of this prediction with experiments
\cite{montrichok03,zeng03} provides  strong justification for the proposed formalism.
Since the spontaneous melting probability is considered as the tunneling phenomena,
there exists a mid point temperature at which the thermal energy becomes equal to the height of the entropic barrier.
This aspect has been utilized further to derive a simple but an important relation between
the double strand stiffness parameter and the mid point temperature which can be determined
experimentally. It is shown that the height and the width of the derivative of the
double strand separation probability with respect to
temperature depend strongly on the parameters
of the entropic barrier and thus, can be used as the experimental tool to learn more
about the nature of the stacking interaction.

\section {Formalism}

The details of the PB model with various improvements have been described elsewhere
\cite{peyrard89,dauxois93,dauxois95,zhang97,nyeo99,dauxois02} and also we have
mentioned it briefly in the appendix for completeness. What concerns here in the
present study is the partition function of the configurational part given by

\equation Z_y=\frac{1}{\alpha^N} \int_{-\infty} ^ \infty \Pi_{i=1}^N  dy_n
             ~e^{-\beta H(y_n,y_{n-1})}, \endequation \label{mzy}

where the constant $\alpha$ having the dimension of length has been
introduced to make $Z_y$ dimensionless.
The potential energy component $H(y_n,y_{n-1})$ is

\begin{eqnarray}  H \left (y_n,y_{n-1} \right )= \sum_n
                                       \frac{1}{2} K \left (y_n-y_{n-1} \right )^2 \nonumber\\
           +  D_n \left [ exp \left (-a_n y_n \right ) -1 \right ]^2 -D_n.
\nonumber \\
          \end{eqnarray} \label{mhy}

The index $n$ runs over all the base pairs and $\sqrt{2} y_n$ is the relative
stretching of the hydrogen bond from the equilibrium position of the $n^{th}$ base
pair. The first term represents the stacking interaction potential between adjacent
base pairs with a $y-$dependent stiffness constant $K$. The Morse potential (second term)
describes the interaction of the base pairs in the two complementary
strands.
The parameters of the model, the depth
$D_n$ and the spatial range $a_n$, distinguish between the A-T and G-C base
pairs.
The integral Eq.(\ref{mzy}) can be evaluated in the thermodynamic limit using the eigen values
and eigen functions of a transfer integral (TI) operator \cite{krumhansl75}

\equation \frac{1}{\alpha}\int dy_{n-1}~ e^{-\beta H(y_n,y_{n-1})}~ \phi(y_{n-1}) =
                       ~e^{-\beta \epsilon}~ \phi(y_n). \endequation \label{mti}

It yields $Z_y=exp(-\beta N \epsilon_0)$ where $\epsilon_0$ is the lowest eigen value of a
Schrodinger type equation

\equation
-\frac{1}{2 \beta^2 K} \frac{\partial ^2 \phi(y)}{\partial y^2}
 +V(y) \phi(y)= (\epsilon + s_0)~ \phi(y), \endequation \label{sch1}

where

\begin{eqnarray} s_0= \epsilon+\frac{1}{2 \beta}~ln~\left (\frac{2 \pi}{\beta K
                 \alpha^2}~\right ). \end{eqnarray} \label{meigen}

As expected, the constant $\alpha$ does not affect the final expression for $Z_y$ or the free
energy per site $f=-(1/N \beta)~ln(Z_y)=\epsilon_0$. However, as will be shown later, it is
an important and useful
parameter of the present formalism which affects the thermal activation energy
$s_0$ and also it is required to make the TI equation (\ref{mti}) dimensionally consistent.
It may be mentioned here that the full partition function that contains all the factors
including the kinetic terms is dimensionless due to the phase space factor $h^{2N}$ in the
denominator. Therefore, the factor $\alpha^N$ in (\ref{mzy}) can be considered as the part
of the phase space distance associated with only the configurational part of the
partition function whose value can be fixed by comparing the results with the experimental
measurements. The factor $K$ is assumed of the form $K=k_0[1+\rho(y)]$ so that
$K$ decreases from $k_0(1+\rho_0)$ to $k_0$
at large $y$. It is now possible to separate
$s_0$ into two parts: a $y-$dependent positive term which gives an entropic barrier when
added to the Morse potential and a $y-$independent thermal part $y_{th}$,
 although both
parts depend on temperature. Accordingly,
Eq.(\ref{sch1}) can be rearranged to give

\equation
-\frac{1}{2 \beta^2 K} \frac{\partial ^2 \phi(y)}{\partial y^2}
 +V_{eff}(y) \phi(y)= (\epsilon + \epsilon_{th})~ \phi(y), \endequation \label{sch2}

where
\equation V_{eff}(y)=\left [ D \left ( e^{-ay}-1 \right )^2 - D \right ]+
                     \frac{1}{2 \beta}~ln \left [1+\rho(y) \right ], \endequation
and
\equation \epsilon_{th} = \frac{1}{2 \beta} ln \left ( \frac{2 \pi}{\beta k_0 \alpha^2} \right ).
\label{eth} \endequation

In the PB model, a y-dependent stiffness with an exponential form $\rho=\rho_0 e^{-b y}$
was used to reflect the fact that the DNA is significantly more rigid in the
double-stranded conformation than when either of the two interacting base pairs
is stretched. Although there is no theoretical guideline,
a value of $b=0.35~{A^o}^{-1}$ was used in \cite{campa98} to simulate the
melting curves
of short heterogeneous DNA sequences (
which accounts for total yield $f$).
However, in this study, we have felt the
need to use a very small value of $b$ in order to explain
the spontaneous melting
fraction of the fully dissociated molecules and partially open
intermediate states simultaneously for many oligomers.
A small value of $b$ would mean a flat entropic
barrier over a large distance. Although it is very much essential for  $\rho$ to
decrease from $\rho_0$ to $0$, the rate of decrease is not found fast enough unless the stretching
exceeds a particular limit which may be due to the
strong backbone support it receives from the neighboring
base pairs while in the double stranded conformation.

\begin{figure}
\centerline{\hbox{
\psfig{figure=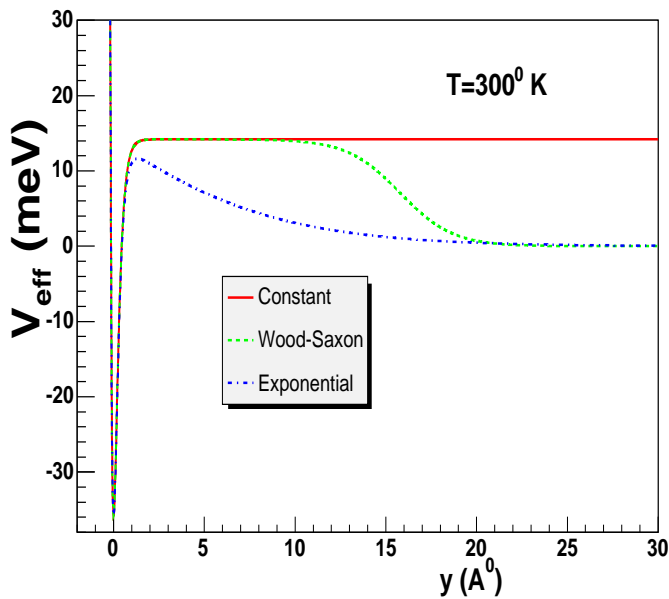,width=4.0in,height=3.6in}}}
\caption{The effective potential $V_{eff}$ (in units of meV )
as a function of $y$.
The parameters are $D=0.05~eV$, $a=4.2~{A^o}^{-1}$, $T=300^0~K$,
$k_0=0.013~eV/{A^o}^2$ and  $\rho_0=2.0$. For exponential dependence, $b$ is
taken as $0.2~{A^o}^{-1}$ whereas for Wood-Saxon form,
$y_0=15~A^o$ and $b=1.42~A^o$.}
\end{figure}

This could be interpreted as some
type of co-operative phenomena involving several base pairs with a
range larger than what is normally expected. To simulate this effect,
we use a slightly different form of Wood-Saxon type

\equation \rho= \frac{\rho_0}{1+e^{-(y-y_0)/b}}, \endequation

characterized by the depth $\rho_0$, range $y_0$ and diffuseness  parameter
$b$.
These are the co-operativity
parameters and, in general, may depend on the sequence composition, length of the sequence and
also on the buffer conditions.
Since it is not known how to fix, these are considered as free parameters
to be fixed by the experimental observations.
Figure 1 shows a typical plot of the effective potential $V_{eff}(y)$ as a
function of
$y$ at a temperature
of $T=300^0 K$. The potential parameters are $D=0.05~ eV$, $a=4.2~{A^o}^{-1}$, $k_0=
0.013~eV/{A^o}^2$ and $\rho_0=2.0$. For exponential dependence $b$ is taken as $0.2~{A^o}^{-1}$
whereas for Wood-Saxon form, $y_0$ and $b$ are chosen as $15~A^o$ and $1.42~A^o$
respectively (as an example).
The solid line corresponds to the case when the y-dependence is ignored.
As can be seen from the figure, although
both the exponential and the Wood-Saxon forms lead
to the same physical consequence, namely, a decrease of the backbone stiffness from
$k_0(1+\rho_0)$ to $k_0$, qualitatively, both give rise to two different forms of
entropic barriers. We have studied both the forms and prefer to retain
the Wood-Saxon form which is found to be
more suitable to explain the melting behavior of several oligomers.

\section {Double Strand Separation Versus Intermediate States}

In the following, we describe the method to characterize the fraction of fully open
molecules $p$ and partially melted intermediate states $\sigma$.
A dsDNA corformation is thermodynamically
stable when the free energy per base pair $\epsilon < 0$. On the otherhand, for ssDNA,
$\epsilon \ge 0$. The bound state solution of the Morse potential with the entropic
contribution with inclusion of both harmonic and anharmonic stacking interaction has
been studied in the past.
For long DNA sequence, the denaturation begins with formation of
bubbles due to local meltings of different domains (local openings of base
pairs) where the base pair bindings are relatively weak. When the average
bubble size attains a critical value at temperature close to the melting point
$(T_m)$,
the entropy gain becomes sufficient to overcome the binding and
the entire sequence separates
into two single strands discontinuously. Therefore, it is only at $T>> T_m$,
a real separation is expected.
However, for shorter sequences as in the case
of many synthetic oligomers, the complete strand separation may occur even
below the melting temperature and the melting process may proceed
with the co-existence of both fully dissociated molecules ($p$)
and partially melted intermediate states ($\sigma$)
over a wide range of temperatures
\cite{montrichok03,zeng03}.
It is noticed that the short oligomers ($< 10$ bp) melt suddenly and the
melting curve can be represented by a two state model in which the molecules are either
completely closed or completely open. As the oligomer size increases, zipper model seems
to be more appropriate which opens continuously giving rise to partially open intermediate
states \cite{zocchi03}. Many synthetic oligomers which are clamped at both the ends by the
relatively GC rich base pairs may nucleate from the middle AT region continuosly until
a limiting bubble size is reached after which the melting becomes discontinuos or even
depending on the sequence composition, the melting may have contributions
from both the processes
simultaneously.
It is also noticed that the oligomers which
are open at one end behave differently than when both the
end points are fixed.
Therefore, the new experimental measurements with different oligomer designs now impose
additional constraints on the theoretical models which are required to
explain both $p$ and $\sigma$ (hence $f$)
simultaneously. Moreover, a quantitative
understanding of the melting process of different synthetic
oligomers will be a stepforward in
gaining more insight of the process by which
the base pairs open up locally in a real DNA sequence even at room temperature.
In this work, we  propose thermally assisted barrier penetration phenomena
as a new mechanism for complete strand separation whereas the partially opened
intermediate states are interpreted as the bound state solutions of the same
effective potential that generates a barrier.
These are two distinct processes  below the melting temperature
although this distinction
vanishes as the temperature is raised up to a limit above which the bound
state no longer exists.
Below the melting temperature, small amplitude coherent fluctuations which involves
several base pairs
may drive a  dsDNA out of the potential well into the continuum
for which $\epsilon \ge 0$. Although thermodynamically favorable to remain
separated, these conformations
are still in the dsDNA state (small amplitude fluctuations around the equilibrium
value $y \sim 0$) and
are required to overcome an entropic barrier (to the right) in order to remain as
stable ssDNA conformation. The fraction $p$, thus, amounts to calculating the
thermally assisted tunneling probability through an entropic barrier with energy $\epsilon_{th}$
as given by (\ref{eth}) where we have set the free energy $\epsilon =0$. The
probability of tunneling can be obtained by solving the Schrodinger Eq.(\ref{sch2})
using WKB approximation

\equation p=\left (1+e^{2S} \right )^{-1}, \endequation
with the action integral

\equation S=\int_{y_1}^{y_2} \left [  2 \beta^2 K(y) \{ V_{eff}(y)-\epsilon_{th} \}
\right ]^{1/2} dy.
\label {p1} \endequation

Note that the term $1/2 \beta^2 K(y)$ is the quantum analog of
$\hbar^2 /2 \mu$. When the thermal energy $\epsilon_{th}$ becomes greater than
the barrier height $V_0$, the potential barrier can be approximated by an inverted
parabola, allowing the use of the analytic expression of Hill and Wheeler \cite{hill53}

\equation p= \left [ 1+exp\{\frac{2 \pi}{\omega}(V_0-\epsilon_{th}) \}\right ]^{-1},
\label{p2} \endequation
with \equation \omega = \left [-\frac{1}{\beta^2 K(y)} \frac{\partial^2 V_{eff}}{{\partial
                         y^2}} \right ]^{1/2},
\endequation
evaluated at the barrier position $y_b$.

Now, using (\ref{p1}) for $\epsilon < V_0$ and (\ref{p2}) for $\epsilon \ge V_0$,
the ssDNA fraction $p$ as a function of $T$ can be estimated. For the intermediate
states, we estimate the quantity

\equation \sigma = N \int_{y_{min}}^{y_{max}} |\phi_0(y)|^2~dy, \label{sig} \endequation

where $N$ is a normalization constant and
$\phi_0(y)$ is the ground state eigen function of the effective potential
$V_{eff}$ for temperature $T \le T_m$.
The integral is bounded by two limits whose values are very crucial
to reproduce the experimental observation of the peak to valley ratio
of the partially melted intermediate states \cite{zeng03}. In general, $y_{min} \ge 0$
so that the
nucleating bubble gains sufficient entropy to grow whereas
$y_{max}$ defines a range
upto which the entropic barrier exists.  It may be mentioned here that
we do not include the contribution of $\phi_0(y)$ beyond $y_{max}$
although $\phi_0(y)$ is still nonvanishing
particularly when $T$ approaches $T_m$. At such high temperature,
the ground state
energy becomes quite close to the continuum and the melting may proceed with opening
of all the remaining base pairs suddenly. Therefore, the fraction of $\phi_0(y)$
beyond $y_{max}$ only contributes to the sudden opening which has been already included in
Eq.(\ref{p2}) for $\epsilon_{th} \ge V_0$. As will be shown later,
the experimental data supports a long entropic
barrier with range $y_0 \sim y_{max}$. Since we are interested for $\phi_0(y)$
in the range $y_{min} \le y \le y_{max}$, where the entropic barrier remains
practically constant,
the ground state solution of the Morse potential
with a constant stiffness $k=k_0(1+\rho_0)$

\equation
\phi_0(y)=\sqrt{a} \frac{(2d)^{d-1/2}}{[\Gamma(2d-1)]^{1/2}} exp(-d e^{-ay})
         exp[(d-\frac{1}{2})a y ]. \label{phi} \endequation

can be used to estimate $\sigma$ (as a good approximation).

It is well known that a sharp melting transition can not be understood only
on the basis of the ground state solution of
Morse potential with a constant stiffness unless a y-dependent stacking interaction
is included. In the present formalism,
the above desired effect is already included in the total yield
$f$ which is now the sum of $\sigma$ obtained with a constant stiffness as part of
the bound state solution and $p$ which
is the tunneling probability through a dynamically generated entropic barrier that
includes stiffness variation.

\begin{figure}
\centerline{\hbox{
\psfig{figure=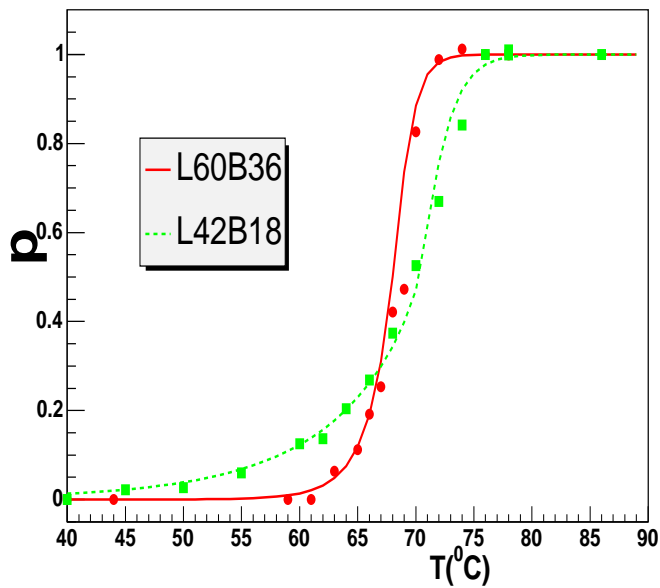,width=4.0in,height=3.6 in}}}
\caption{The fraction of open molecules $p$ as a function of
temperature $T$ (in $^{0}C$).
The parameters are $D=0.0494~eV$, $a=4.2~{A^o}^{-1}$,
$k_0=0.013~eV/{A^o}^2$, $\rho_0=1.99$,
$y_0=40~A^0$, $b=4.6~A^0$ and $\alpha=2.18~A^o$ for
$L60B36$ sequence and
$D=0.0505~eV$, $a=4.2~{A^o}^{-1}$,
$k_0=0.013~eV/{A^o}^2$, $\rho_0=2.01$,
$y_0=15~A^0$, $b=1.42~A^0$ and $\alpha=2.18~A^o$ for
$L42B18$ sequence. The filled circles and squares are experimental
data points}
\end{figure}

Now using the above formalism, we have estimated both the fraction $p$ and
$\sigma$ for two synthetic oligomers L60B36 and L42B18 which are
studied experimentally in ref\cite{zeng03}. Both the oligomers are clamped at the ends
with identical GC pairs and with an AT rich middle region of variable lengths (L60B36: length L=
60 bases, bubble forming region B=36 bases, L42B18: L=42 bases, B=18 bases). Figures
2 and 3 show the estimated fractions of completely separated molecules $(p)$ and partially melted
intermediate states $\sigma$ as a function of temperature $T$.
The parameters used in the calculations are listed in the figure captions.
The agreement of
the theoretical predictions with the experimental measurements is indeed excellent.

\begin{figure}
\centerline{\hbox{
\psfig{figure=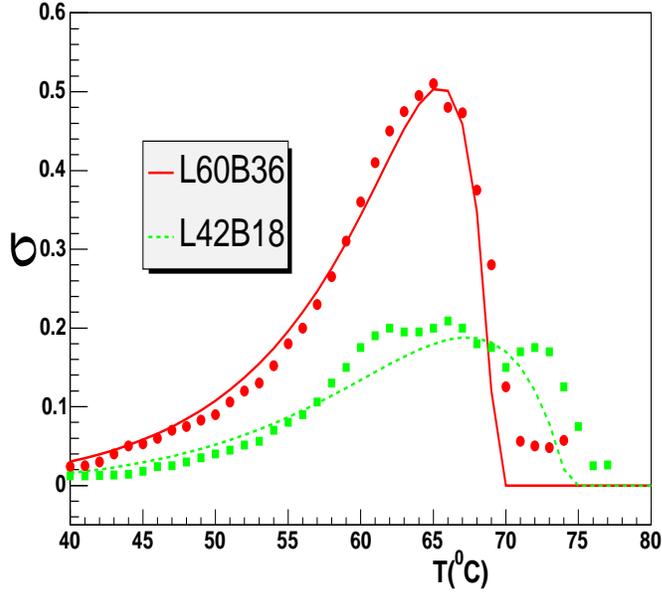,width=4.0in,height=3.6in}}}
\caption{The quantity $\sigma=f-p$ as a function of
temperature $T$ (in $^{0}C$).
The parameters are  $D=0.0494~eV$, $a=4.2~{A^o}^{-1}$, $k_0=0.013~eV/{A^o}^2$,
$\rho_0=1.99$,
$y_{min}=9~A^0$ and $y_{max}=40~A^0$  for
$L60B36$ sequence and
$D=0.0505~eV$, $a=4.2~{A^o}^{-1}$, $k_0=0.013~eV/{A^o}^2$,
$\rho_0=2.01$,
$y_{min}=9~A^0$ and $y_{max}=15~A^0$ for
$L42B18$ sequence. The normalization constant $N$ is taken as unity.
The filled circles and squares are experimental data
points.}
\end{figure}

Although, initially
there are too many free parameters, most of them have been fixed based on different
physical considerations. For
example, it is noticed that the tunneling probability $p$ is not very sensitive
to the Morse potential as long as they are reasonable,
but depends sensitively on the factor $K(y)$ that decides the nature
of the entropic barrier. On the otherhand, the fraction $\sigma$ and the peak position
have strong dependence on the depth and range of the Morse potential.
As will be shown later,we fix the Morse parameters that give optimum
fits to the $\sigma$ measurements. Accordingly, we take
$D=0.0494~eV$ for L60B36 and $D=0.0505~eV$ for L42B18 whereas $a$ is fixed at $4.2~{A^o}^{-1}$
for both the sequences. In fact, these values are nearly identical to what has been used
in \cite{campa98,kalosakas03} corresponding to an AT base pair. As $p$ is
the tunneling probability, there exists a temperature $T_d$ at which the thermal activated
energy $\epsilon_{th}$ becomes equal to the barrier height $V_0$
(tunneling probability becomes $0.5$).
Since the contribution of the Morse potential is  negligible
outside the well, it is basically the flat portion of the stacking interaction
$1/(2 \beta) ln(1+\rho_0)$ that decides the height of the
entropic barrier. Equating $\epsilon_{th}$ with
$V_0$ at $T_d$ and using the approximation $V_0 \approx ln(1+\rho_0)/(2 \beta)$, we get

\equation T_d = \frac{\alpha^2 k_0 (1+\rho_0)}{2 \pi k_\beta} \label{td}. \endequation

Thus, the above relation puts the constraint on the choice of $k=k_0(1+\rho_0)$ and $\alpha$.
For same $T_d$,  we may have different combinations of $k$ and $\alpha$. In
practice, $k$ is
really not a free parameter which can be varied randomly as it also controls the temperature
dependent effective
mass that appears in Eq.(\ref{sch2}).  Although, we have tried various combinations,
it is found that the fit is optimum when $\alpha$ is close to $2 A^o$.
This is an interesting observation as this distance also corresponds to an average
stretching beyond which the hydrogen bonds start breaking.
As we have discussed before, $\alpha$ is a
phase space factor associated with the configurational partition function. It
is probably meaningful to expect $\alpha$ of the order of $2 A^o$ for the
configurational changes to occur. Subsequently, we fix $\alpha$ at
$2.18 A^o$ which is an
optimum choice  for
all the sequences. With $\alpha$ fixed,
Eq.(\ref{td}) now gives a very simple but an
important relation which can be used to estimate the stiffness parameter $k=k_0(1+\rho_0)$
directly by knowing $T_d$ which can be measured experimentally. For example,
from the experimental curves,
we estimate $T_d \sim 68.1^o C$ for L60B36 and $T_d \sim 70.36^o C$ for L42B18. With
the choice of $\alpha=2.18 A^o$, we get $k \sim 0.0389~eV/{A^o}^2$ and $k \sim
0.391~eV/{A^o}^2$ for the
above two sequences. Although, knowing the full stiffness $k$ is itself important,
yet we do not know how to split it between $k_0$ and $\rho_0$. However, taking the guidance
from ref \cite{campa98} where $\rho_0$ was taken  $\sim 2$, we find $k_0$ should be  $\sim 0.013~
eV/{A^o}^2$. Subsequently, we fix $k_0$ at $0.013~eV/{A^o}^2$ for all the sequences and
consider the co-operative parameter $\rho_0$ as a variable which, although should be close to
$2$, can be used as tuning parameter so as to reproduce the correct mid point temperature
$T_d$. Thus, with $D$, $a$, $\alpha$, $k_0$ and $\rho_0$ fixed,
we are now left with only two free parameters $y_0$ and $b$ which are finally
adjusted to reproduce the experimental measurements. As listed in the figure captions,
the strength $\rho_0$ turns out to be $1.99$ for L60B36 whereas it is $2.01$ for L42B18.
This slight variation is expected as both the sequences have different dissociation
temperature $T_d$. Similarly, the range and diffuseness parameters are found to be $40 A^o$
and $4.6 A^o$ for L60B36 sequence. The corresponding values for L42B18 are $15 A^o$ and
$1.42 A^o$ respectively. Although it is difficult to draw any definite conclusion,
the length $y_0$ seems to depend strongly on the size of the bubble forming
AT rich region or
more precisely on the ratio $B/L$ which are $0.6$ and $0.43$ respectively
for the above two oligomers. It may be
mentioned here that this observation is more
meaningful for these two oligomers which have identical endpoints and may not be
valid in general.  However, all sequences studied in this work
 have a nearly flat entropic barrier
over a larger range.  This effect, probably,  can be interpreted as the
reminiscent of a long range co-operative phenomena.

The parameters for $\sigma$ are also fixed in a similar way.
Since the entropic barrier turns out to be
nearly constant up to the range $y_0$,
$\sigma$ has been estimated
using Eq.(\ref{sig}) with the ground state wave function
given by Eq.(\ref{phi}). It is
noticed that the peak position of the yield is very sensitive to the
melting temperature $T_m=2\sqrt{2kD}/(k_\beta a)$ \footnote{Note the distinction
between $T_d$ and $T_m$ where $T_d$ is the dissociation temperature at which
$p$ becomes half whereas $T_m$ is the melting temperature above which the
bound state vanishes. In general, $T_d$ and $T_m$ need not coincide.}.
For a given $k$ which is
fixed from the $p$ yield, $\sqrt{D}/a$ ratio (the $p$ yield is not sensitive to
this ratio), is adjusted to have the correct peak position whereas
the limit of integrations $y_{min}$ and $y_{max}$ are adjusted to reproduce
the observed peak to valley ratio. As discussed before,
we take $a=4.2~{A^o}^{-1}$ as in \cite{campa98} and
use $D=0.0494~eV$ for L60B36 and $D=0.0505~eV$ for L42B18 to reproduce the $\sigma$
curves. It is also
noticed that the optimum results are obtained when $y_{max} \sim y_0$, i.e.
the range of the entropic barrier. Subsequently, we set $y_{max} = y_0$ and adjust
$y_{min}$ which turns out to be $9 A^o$ for both the sequences.
Therefore,  by putting the constraint
to explain both $p$ and $\sigma$ measurements simultaneously, we have fixed
most of the free parameters of the model. Although, $y_{min}$ is expected to be
$\ge 0$, a value of $9 A^o$ is indeed too large as compared to the minimum threshold
of average stretching $<y> = 2 A^o$ above which hydrogen bond between the base pairs
starts breaking. This large value of $y_{min}$ probably is very specific to these two
oligomers whose end points are fixed by a relatively stronger GC rich regions. An unzipping
transition leading to the continuous melting may begin with a nucleating bubble initiating
from the softer AT rich middle region. Since this melting is a competition between the
energy of the bound states and the entropy of the dissociated states, the transition
may not proceed until the bubble radius attains a minimum value which
is about $9 A^o$ for these oligomers.
A stretching distance
of $y_{min} \sim 9 A^o$ would mean a nucleating bubble of diameter $d \sim 18 A^o$
containing about $4$ to $5$ open base pairs. Unless the  size
attains this minimum value, the entropy
gain is not sufficient enough for the bubble to grow. Note that
this minimum size is different from the critical bubble size which is required
to drive a spontaneous transition. The $y_{min}$ value
strongly depends on the size and composition of the sequence. We have found
that $y_{min}$ value is much smaller if the sequence is  homogeneous. Since
the bubble forming AT region of
L60B36 and L42B18 oligomers are
bounded by relatively stronger GC rich regions at the end, the denaturated
bubble may encounter higher resistance to grow due to strong binding. This
is to say that
the interaction of the bubble with
itself and also with its surroundings
may bring a reduction in entropy. Therefore,
for a given entropy gain, the minimum bubble size
required to grow (by compensating the binding energy)
for the sequence with fixed end points
is larger then the case
when the sequence is open at both the ends.
This could be
viewed as analogous to the excluded volume interaction whose effect is to
reduce the entropy of the loop. Notice that
when $y_{min}$ is large or the range
$(y_{max}-y_{min})$ is small, the $\sigma$ contribution decreases.
Another parameter which may also  suppress $\sigma$
yield is the strong binding effect of the base pair. If
the binding energy is large, i.e. if the $\sqrt{D}/a$ ratio is large,
the temperature $T_m$ at which the bound state vanishes goes up where
as the dissociation temperature $T_d$ remains unaffected. If $T_d >> T_m$,
complete dissociation may take place much earlier,
thus reducing the fraction of continuous melting.
Therefore, in either case, the most
probable route
available for denaturation is the sudden melting thus making the transition
to appear discontinuous.

It may be mentioned here that for a complete strand separation, it is not
necessary to raise the temperature until all the base pairs break. Normally,
the nucleation is initiated with a small denaturating bubble from the
softer AT rich region. When the average bubble size attains a critical value,
the loop entropy becomes sufficient enough to overcome the binding so that
rest of the base pairs unwind instantly. The critical sized bubble is
formed either through a sequential unzipping process what is known as a
continuous melting  or through the sudden opening due to coherent fluctuation
(followed by tunneling). In the first approach, the critical size is
reached when the temperature is close to the melting point  whereas the
second process can happen at any temperature with different probabilities. Since
the tunneling probability $p$ depends only on the entropic barrier
parameters, the heterogeneity of the sequence which introduces a site
dependent binding $V_n$, does not affect the full dissociation yield. On the
otherhand, the fraction of the continuous yield $\sigma$ is strongly
affected by the heterogeneity of the sequence. By design, the bubble formation
for L60B36 and L42B18 is initiated in the AT rich middle region. Since the
AT rich region for L60B36 is long enough, the critical bubble size may
still remain confined to the homogeneous middle portion only.  For L42B18, the
homogeneous AT region may not be long enough and the bubble may extend
upto the GC rich region also. As a consequence, $\sigma$ transition may
encounter different melting regions with appropriate  weights.
As can be seen, the
$\sigma$ peak for L60B36 is more sharper as compared to the case of
L42B18 sequence where the GC region may start contributing. This
curve can be better understood as a weighted sum over
two melting regions with appropriate weights although $p$ curve can still be
understood as a single tunneling phenomenon
through an appropriate entropic
barrier.

Figures 4 and 5 show the comparison of
$f$, $p$ and $\sigma$ with the corresponding experimental results using
the set of parameters as listed in the figure captions 2 and 3. The linear rise of
the total open base pairs $f$ after the
sigmoidal transition is a well-known phenomenon
attributed to the residual base unstacking in the single
strands.

\begin{figure}
\centerline{\hbox{
\psfig{figure=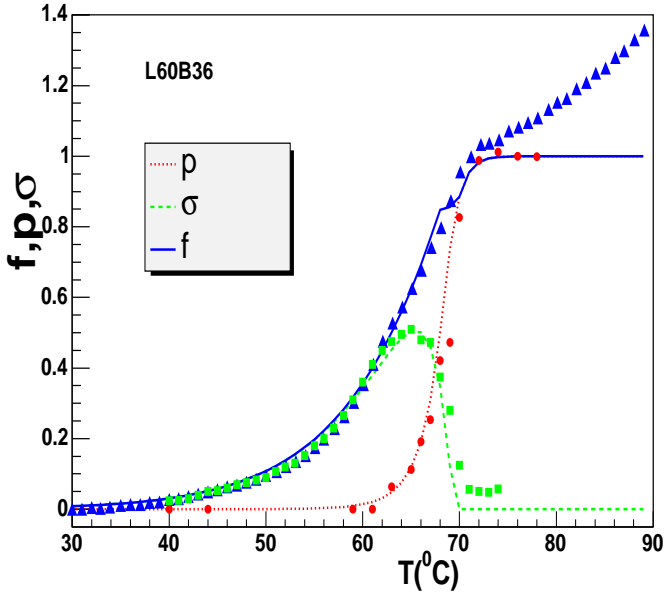,width=4.0in,height=3.6in}}}
\caption{The quantities $f$, $p$ and $\sigma$ as a function of
temperature $T$ (in $^{0}C$) for
$L60B36$ sequence.
The filled triangles, circles and squares are the
corresponding experimental data points.}
\end{figure}

However, the theoretical
prediction of $f$ does not increase beyond unity as this effect has
not been included in the present model.

As shown in figures (2-5), the present model has been quite successful
in explaining both qualitatively and quantitatively experimental
observations.
We have also studied two more oligomers L42V1 and L48AS
for which experimental data points are available \cite{zocchi03}. The
L42V1 sequence is $42$ base pairs long with $18$ AT base pairs in the
middle whereas L48AS is $48$ base pairs long with a AT rich region of
$21$ base pairs which is open at one end.

Figure 6 shows the plot
of $p$ as a function of $T$ for the above two oligomers with the parameters
listed in the figure captions. The $y_0$ parameter is found out to be
$18~A^o$ for L42V1 with a diffuseness value of $2.2~A^o$. This oligomer
(a $B/L$ ratio of $\sim 0.42$) is quite identical to L42B18 oligomer in design.
Accordingly, the parameters of $\rho(y)$
follows the systematic and also the full dissociation probability is
quite close to that of L42B18 oligomer.

\begin{figure}
\centerline{\hbox{
\psfig{figure=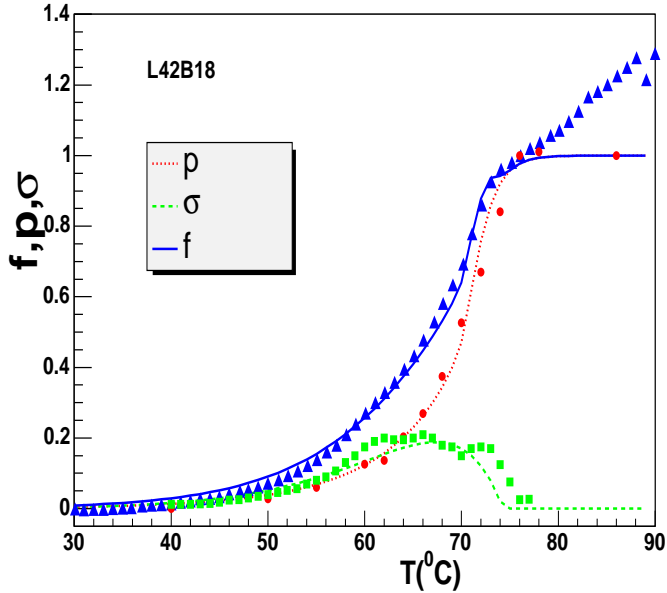,width=4.0in,height=3.6in}}}
\caption{The quantities $f$, $p$ and $\sigma$ as a function of
temperature $T$ (in $^{0}C$) for
$L42B18$ sequence. The filled triangles, circles and squares are the
corresponding experimental
data points.}
\end{figure}

Interestingly, this oligomer
has no contribution to $\sigma$ as it melts suddenly at all temperatures.
As mentioned before, the vanishing of $\sigma$ component would mean
a vanishing range ($y_{max}-y_{min}$)  or
a higher effective binding energy which pushes the melting temperature
$T_m$ to the higher side. Since L42V1 is quite similar to L42B18 in design,
it is difficult to understand why $\sigma$ vanishes for the former whereas
this component is still significant for the later oligomer. Although this
needs more careful analyses, a possible reason could be different buffer
conditions in which the DNA solutions were prepared. It is known that a
stronger ionic concentration may increase the binding (higher $D$, hence
higher $T_m$) while the entropic parameters may still remain unchanged.
On the otherhand, the melting behavior of L48AS is quite
different from the rest of the oligomers as it opens from one end. It has
a range which is relatively larger $(y_0 \sim 60~A^o$ and $b=9~A^o$).
The large entropic range
explains why $p$ component is small with a relatively larger $\sigma$
contribution. We have not estimated the $\sigma$ fraction, but it is
expected to be broad with contributions coming from two
different melting regions.

\begin{figure}
\centerline{\hbox{
\psfig{figure=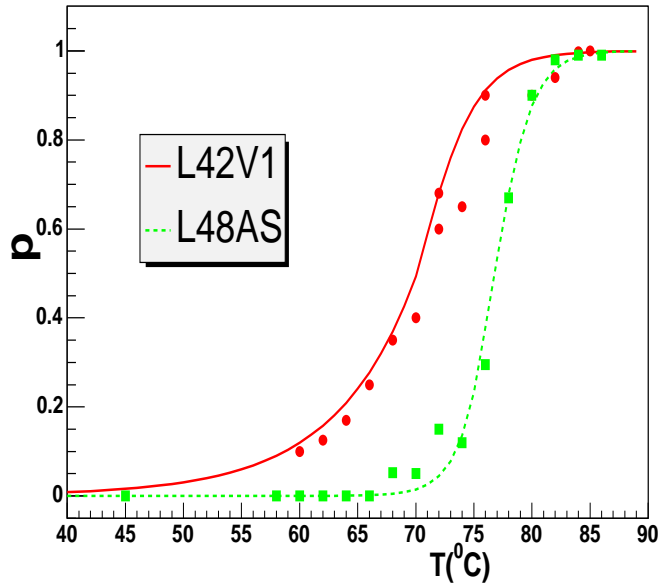,width=4.0in,height=3.6in}}}
\caption{The fraction of open molecules $p$ as a function of
temperature $T$ (in $^{0}C$).
The parameters are $D=0.0505~eV$, $a=4.2~{A^o}^{-1}$,
$k_0=0.013~eV/{A^o}^2$, $\rho_0=2.01$,
$y_0=18~A^0$, $b=2.2~A^0$ and $\alpha=2.18~A^o$ for
$L42V1$ sequence and
$D=0.0505~eV$, $a=4.2~{A^o}^{-1}$,
$k_0=0.013~eV/{A^o}^2$, $\rho_0=2.07$,
$y_0=60~A^0$, $b=9~A^0$ and $\alpha=2.18~A^o$ for
$L48AS$ sequence. The filled circles and squares are experimental
data points}
\end{figure}

The temperature driven denaturation has also been studied for
synthetic  ploy (A)- ploy (T)
oligomers of various lengths in different buffer conditions \cite{zocchi03}. These oligomers
can be regarded as homogeneous duplexes held together by identical base pair interaction
at the different sites. Although, the measurements do not separate the fractions
into fully dissociated and partially open components, the total fraction $f$ is
itself interesting as it has a large not so well understood premelting region.
The peak and width of the derivative  with respect to temperature also show strong dependence
on the length of the oligomer as well as on the buffer conditions.
Due to finite size effects, the co-existence of both the components is expected
for these oligomers also.
Since data is not available, we will not try to estimate $p$ and $\sigma$ exactly
so as to reproduce
the total yields, rather we will discuss a scenario based on which the experimental results
can be understood at least qualitatively. Again as before, we consider an arbitrary oligomer
which may have similar potential parameters like L60B36 oligomer except for the fact that
 this oligomer is now homogeneous. A homogeneous sequence may have a lower $y_{min}$ value
than when the sequence composition is more heterogeneous.

\begin{figure}
\centerline{\hbox{
\psfig{figure=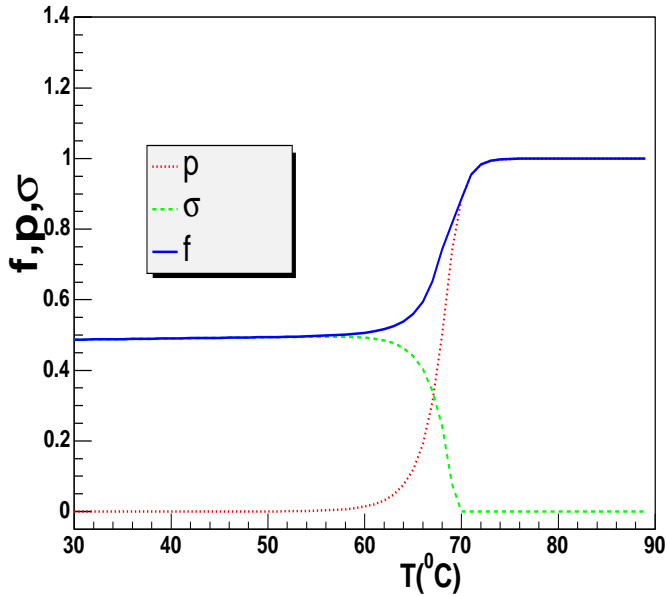,width=4.0in,height=3.6in}}}
\caption{The quantities $f$, $p$ and $\sigma$ as a function of
temperature $T$ (in $^{0}C$) for
an arbitrary sequence.
The parameters are $D=0.0494~eV$, $a=4.2~{A^o}^{-1}$,
$k_0=0.013~eV/{A^o}^2$, $\rho_0=1.99$,
$y_0=40~A^0$, $b=4.6~A^0$, $\alpha=2.18~A^o$,
$y_{min}=0$, $y_{max}=y_0$ and $N=0.5$.
}
\end{figure}

As a model study, we choose
$y_{min} =0$, i.e. the minimum expected value. As the $y_{min}$ decreases, the behavior of
$\sigma$ changes completely, becoming more uniform as $y_{min}$ decreases from a higher
threshold value towards $0$. Figure 7 shows a typical plots of $\sigma$, $p$ and  $f$
as a function of $T$ where we have set $y_{min}$ to its lowest value. Although
we have used a
normalization factor $N=0.5$ , it can be adjusted so as to reproduce
the appropriate yield in the premelting region.
The total yield is quite similar (at least qualitatively)
to what one observes experimentally
\cite{zocchi03}. Interestingly, the so called premelting region is probably due to
a continuous transition which has nearly uniform
probability up to temperature $T_m$ above
which the transition becomes discontinuos. Since below $T_m$, the continuous
process dominates, a zipper type model may be found more appropriate to
describe this type of melting curve, particularly when the sequence is
homogeneous and also relatively large. It is also observed that the length
of the entropic range $y_0$ depends on the length of the
sequence, i.e. for the same composition, $y_0$ increases with the length
of the sequence probably until an upper limit is reached. So the smaller
sequence will have short ranged entropic barrier with increased $p$
contribution. When the size is very short
($< 10$ bp), it may completely dissociate without any intermediate state.
This explain why a two step model is found more appropriate to describe the
melting behavior of very short oligomers.

Since the premelting region is flat, the non vanishing part
of the derivative of the total
yield is basically due to the fraction $p$. Figure 8 shows the plot of
$dp/dT$ (estimated numerically) as a function of $T$ for different co-operativity
parameters.
The curves on then left correspond to the case when $\rho_0=1.8$ whereas
the curves on the right are for $\rho_0=1.99$. On both the sides, the solid
curve corresponds to the parameters as listed in the figure captions while
the dotted and dashed curves are obtained when either $y_0$ or $b$ is changed. The point
which needs to be stressed here is that the peak positions (hence the temperature $T_d$)
are quite sensitive to the strength $\rho_0$ which may be small for smaller oligomers.
Therefore, the shift in peak positions of the oligomers  to the higher values when the
length increases can be understood on the basis of the increase of the $\rho_0$
parameter. On the otherhand, the width and the height are very sensitive to $y_0$ and $b$
parameters which also depend on the length of the oligomers as well as on the buffer
conditions. Therefore, with suitable $\rho_0$, $y_0$ and $b$ parameters, it is possible
to understand all aspects of the melting behavior of the synthetic oligomers studied
in \cite{zocchi03}.

\begin{figure}
\centerline{\hbox{
\psfig{figure=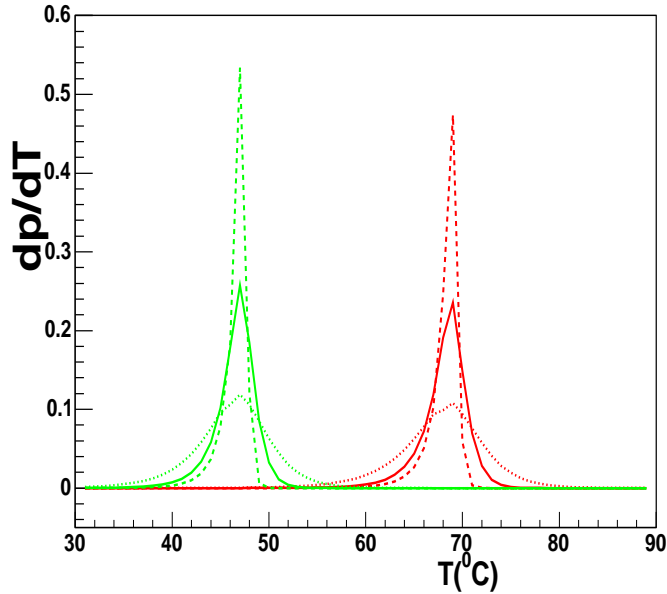,width=4.0in,height=3.6in}}}
\caption{The derivative of $p$ as a function of
temperature $T$ (in $^{0}C$) for some arbitrary
sequence.
The parameters are $D=0.0494~eV$, $a=4.2~{A^o}^{-1}$,
$k_0=0.013~eV/{A^o}^{2}$, $\alpha=2.18~A^0$,
$y_0=40~A^0$ and $b=4.6~A^0$.  The curves on the left
correspond to a $\rho_0$ value of $1.8$ whereas the curves
on the right are for $\rho_0=1.99$. In both sets of curves, the solid lines
correspond to the parameters as listed above. The dotted lines are
obtained when only $y_0$ is changed from $40~A^o$ to $32~A^o$
 and the dashed lines
are obtained when $b$ is changed from $4.6~A^o$ to $3.5~A^o$.}
\end{figure}

\section {Conclusions}
Based on the above studies, we have several conclusions to make.
Recall again, a large DNA sequence
melts with the initiation of the nucleating bubbles at different regions which are
relatively soft.  A complete strand separation occurs when the average bubble
size attains a critical value. Depending on the nature of the
entropic barrier, the domain melting occurs either through continuous opening
of base pairs or due to coherent fluctuation followed by tunneling.
However, at this level,
whatever may be the process of domain melting, the complete dissociation of the full
sequence will not be realized
unless the average bubble size attains a limiting value so that the entropic
gain becomes sufficient to overcome the bound state energy. Once the melting
point is reached, a complete
strand separation occurs what appears as a discontinuous phase transition. Therefore, the
complete DNA denaturation in the thermodynamic limit  can be interpreted as a first
order phase transition characterized by a sharp melting temperature at which the
bound state vanishes. However, due to finite size effect as in the case of oligomers,
both type of processes, one due to continuous opening and the other due to the sudden
melting, may co-exist and the concept of a phase transition is not strictly applicable.
Therefore, we invoke a barrier penetration phenomena along with the bound state solution
to explain these two  fractions separately. Although, it is not possible
to draw any definite conclusions unless more experimental data sets of various
synthetic oiligomers are
available with different design, we can only make a few qualitative remarks as
follows.
The available measurements seem to support a long ranged entropic barrier
with parameters which depend on the length and composition of the sequence.
For the same length, the range of the sequence made up of GC base pairs
is higher than when the
sequence is homogeneous with AT base pairs.
 Consequently, the GC rich sequence will have less $p$
and more $\sigma$ contributions as compared to the AT rich sequence.
If the oligomer is short enough ($< 10$ bp) and
homogeneous, a sudden melting may occur due to an entropic barrier which is very
short ranged (small value of $y_0$). These oligomers will also not have
any continuous fraction due to vanishing $(y_{max}-y_{min})$ range and the full melting
process can be explained by the so called
two state model. On the otherhand, with increasing size of the oligomers, the entropic barrier
range increases (due to stronger collectivity)  and the full dissociation probability $p$
decreases. If $y_{min}$ is close to zero, continuous opening takes place with equal
probability at all temperatures until the melting point is reached. As seen in
\cite{zocchi03}, these type of
processes are characterized by a slowly rising pre-melting region followed by a sudden
transition and are more suitable to be described by the zipper type model as contrast
to the two state phenomena. The melting behavior of the oligomers studied in \cite{zeng03}
can also be understood as the sum of two components. These two oligomers L60B36 and L42B18
with the $B/L$ ratios of $0.6$ and $0.43$ are clamped at both the ends with identical
GC rich regions. Since the relative AT length is large enough, the
bubble forming region is homogeneous and the corresponding
$y_0$ values of $40 A^0$
and $15 A^0$ are probably following the $B/L$ (relative length)
systematic. The L60B36 has a longer entropic
barrier than L42B18 and  the corresponding fraction $p$ is also less for L60B36. On the
otherhand, the entropic range is smaller for L42B18. Accordingly,
the $\sigma$ contribution is also
less as compared to L60B36. However, the key aspect where these two oligomers differ
from not being homogeneous is the higher $y_{min}$ value of about $9 A^0$ which makes
the peak to valley ratio in the $\sigma$ yield very distinct. The oligomer
L42V1 which is nearly identical to L42B18
has also been studied in \cite{montrichok03}. While this oligomer
has a comparable $p$ yield, it does not have
any $\sigma$ component. This aspect is interesting and needs further
investigations.
On the otherhand, the oligomer L48AS has a $B/L$ ratio of $0.25$ with more
GC components and also with a large entropic range. Consequently, the
$p$ component for this type of oligomer
is negligible and the transition mostly proceeds through continuous opening. For the same
reason, as remarked in \cite{zeng03}, even shorter oligomer (down to $19$ bp) which
unzips from one end may open continuously.

Finally, we would like to add here that the shape of the intermediate fraction yield $\sigma$
 depends strongly
on the sequence heterogeneity since $V_n$ becomes site dependent. However, the full dissociation
probability $p$ is still can be obtained from the barrier penetration formalism as it is
sensitive only to the entropic barrier. The temperature at which the full dissociation probability
becomes half is directly related to the DNA stiffness parameter. For the sequence for which $T_d=T_m$,
the mid point of the full absorption curve $f$ can be used to estimate the stiffness parameter.
In general, $T_d$ need not be same as $T_m$. Since $T_d$ relates directly
to the stiffness parameter and also its derivative is more sensitive
to the parameters of the entropic barrier,
 it will be more meaningful to study these two
components $p$ and $\sigma$ separately.

\acknowledgements
We would like to thank Prof. Giovanni Zocchi and Yan Zeng for providing
the tabulated values of the experimental data points.

\appendix
\setcounter {figure}{0}
\renewcommand{\thefigure}{A\arabic{figure}}
\section *{A}

A linear chain of DNA molecules can be described by the Hamiltonian given by

\equation H=\sum_n \frac{1}{2} m (\dot u^2_n+\dot v^2_n)+\frac{1}{2}K
           \left [(u_n-u_{n-1})^2+(v_n-v_{n-1})^2\right ]+V(u_n-v_n),
          \endequation \label{ham1}

where $u_n$ and $v_n$ correspond to the displacements of the bases at
site $n$ along the direction of the hydrogen bond, and $V$ is the potential
between the base pairs which can be of the Morse type

\equation V(u_n-v_n)= D \left [ exp \left \{-A(u_n-v_n) \right \}-1 \right ]^2.
\endequation \label{morse1}

In terms of the variables $x_n=(u_n+v_n)/\sqrt{2}$ and $y_n=(u_n-v_n)/\sqrt{2}$,
the above Hamiltonian can be written as
\equation H=\sum_n \left \{ \frac{p_n^2}{2m}+\frac{q_n^2}{2m} \right \}
            + H \left (x_n,x_{n-1} \right ) + H \left (y_n,y_{n-1} \right ),
           \endequation \label{ham2}
where $p_n=m \dot x_n$ and $q_n = m \dot y_n$ are the canonical momenta, and

\equation  H \left (x_n,x_{n-1} \right )= \sum_n \frac{1}{2} K \left (x_n-x_{n-1} \right )^2,
           \endequation \label{hx}

\equation  H \left ( y_n,y_{n-1} \right )= \sum_n \left \{
                                       \frac{1}{2} K \left (y_n-y_{n-1} \right )^2
                    +  D \left [ exp \left (-A \sqrt{2} y_n \right ) -1 \right ]^2 \right \}.
           \endequation \label{hy}

The statistical mechanics of the model is described by the partition function

\equation Z= \frac{1}{h^{2N}} \int_{-\infty} ^ \infty \Pi_{i=1}^N~ dx_n dy_n dp_n dq_n
             ~e^{-\beta H(p_n,x_n,q_n,y_n)}, \endequation

where $\beta=1/(k_\beta T)$ with $k_\beta$ being Boltzmann's constant, and
$h$ is Plank's constant. The integrals for the variables $p_n$, $q_n$, and $x_n$
are of the Gaussian type, and give, respectively
\equation Z_p=Z_q=\left (2\pi m k_\beta T \right)^{N/2},~ Z_x=\left ( 2 \pi
k_\beta T /K \right )
             ^{N/2}. \endequation
The remaining configurational partition function $Z_y$ involves nonlinear Morse potential,
and given by

\equation Z_y= \frac{1}{\alpha^N} \int_{-\infty} ^ \infty \Pi_{i=1}^N~  dy_n
             ~e^{-\beta H(y_n,y_{n-1})}. \endequation \label{zy}

The factor $\alpha$ (dimension of length) has been in introduced to make
$Z_y$ dimensionless. The above integral can be evaluated
by using the transfer integral (TI) method defined by
\equation \frac{1}{\alpha}\int dy_{n-1}~ e^{-\beta H(y_n,y_{n-1})}~ \phi(y_{n-1}) =
                       ~e^{-\beta \epsilon}~ \phi(y_n). \endequation \label{ti}

To compute the integral, we consider a finite chain of $N$ base pairs and chose periodic
boundary conditions by demanding that $y_N=y_0$. This amounts to adding a fictitious
base pair with index $0$ which has the same dynamics as base pair $N$, with a simultaneous
introduction of a factor $\delta(y_N-y_0)$. Consequently, Eq.(\ref{zy}) can be written
as

\equation Z_y=\frac{1}{\alpha^N}
               \int \Pi_{n=0}^N ~dy_n~ \left [ \Pi_{n=1}^N
             e^{-\beta H(y_n,y_{n-1})} \right ]~ \delta(y_N-y_0). \endequation \label{zy1}

The $\delta$ function of $Z_y$ can be expanded on the base of the eigenfunctions of the
TI operator as $\delta(y_N-y_0)=\sum_i~\phi_i^{*}(y_N)~\phi_i(y_0)$. Performing the
integrals over $y_0$, $y_1$, $y_2$,....,$y_{N-1}$ successively yields

\equation Z_y~=~\sum_i~e^{-\beta N \epsilon_i}~\int dy_N~ |\phi_i(y_N)|^2~=~\sum_i e^{-\beta N \epsilon_i}. \endequation
In the thermodynamic limit $(N \rightarrow \infty)$, the result is dominated by the
term having the smallest value of $\epsilon_i$ which we denote by $\epsilon_0$.
The free energy per site is therefore
\equation f=-\frac{k_\beta T}{N} ln(Z_y) = \epsilon_0. \endequation
More interesting for the study of DNA denaturation is the mean stretching $<y>$ of
the hydrogen bonds. It is given by
\equation <y>~=~\frac{1}{Z_y} \int \Pi_{n=1}^N~y~e^{-\beta H(y_n,y_{n-1})} dy_n. \endequation
The integral can again be calculated with TI method and yields

\equation <y>~=~\frac{\sum_{i=1}^N~<\phi_i(y)|y|\phi_i(y)>~e^{-\beta N \epsilon_i}}
 {\sum_{i=1}^N~<\phi_i(y)|\phi_i(y)>~e^{-\beta N \epsilon_i}}. \endequation

Again, in the limit of large $N$ and considering only the lowest eigen value
$\epsilon_0$, the average stretching is given by
\equation <y>~=~<\phi_0(y)|y|\phi_0(y)>~=\int |\phi_0(y)|^2 dy, \endequation
for the normalized eigenfunction $\phi_0(y)$. Note that with the transformation of
the co-ordinates, $\sqrt{2}<y>$ gives the mean stretching of the hydrogen bond
between the two opposite base pairs.

The problem is therefore to find the eigenfunctions and eigen values of the TI operator.
The calculation can be made analytically in the limit of strong coupling between sites.
For large $K$, when $y_n$ differs significantly from $y_{n-1}$, the term $K(y_n-y_{n-1})^2$
grows rapidly, so that the TI is dominated by the value of $y_{n-1}$ which are close
to $y_n$. By defining $y_{n-1}=y_n+z$ and expanding $\phi(y_n+z)$ in powers of $z$,
Eq.(\ref{ti}) can be written as

\equation \frac{1}{\alpha} \int_{-\infty} ^ \infty e^{\frac{-\beta}{2} K z^2}
           ~\left [ \phi(y)+z \frac{
              \partial \phi(y)}{\partial y} + \frac{1}{2} z^2 \frac{\partial^2 \phi(y)}
              {\partial y^2} ~\right ]
            = e^{-\beta [\epsilon -V(\sqrt{2} y)]}~ \phi(y),
         \endequation
where the index $n$ has been dropped. Since the odd term of the Gaussian integral in
$z$ vanishes, the above integral reduces to
\equation
\sqrt{\frac{2 \pi}{\beta K \alpha^2}} ~
      \left [ \phi(y) + \frac{1}{2 \beta K}  \frac{\partial ^2
         \phi(y)}{\partial y^2} ~\right ] = e^{-\beta [\epsilon -V(\sqrt{2} y) ]}~ \phi(y).
            \endequation
The factor $\sqrt{\frac{2 \pi}{\beta K \alpha^2}}$ can be absorbed in the eigen value
by redefining
\equation \bar \epsilon= \epsilon+\frac{1}{2 \beta}~ln~\left (\frac{2 \pi}{\beta K
                 \alpha^2}~\right ). \endequation \label{eigen}

In many practical situations, the relevant magnitude of ($\epsilon-V) \sim D$, the depth
of the Morse potential.
For positive $y$ (as in the case for stretching), the Morse potential is bounded by
$D$. If $\beta D < 1$, the exponential can be expanded to get Schrodinger type
equation
\equation
-\frac{1}{2 \beta^2 K} \frac{\partial ^2 \phi(y)}{\partial y^2}
 +V(y) \phi(y)= \bar \epsilon~ \phi(y), \endequation
where
\equation V(y)=D \left [ e^{-a y} -1 \right ]^2,~~a=\sqrt{2} A, \endequation
and $\bar \epsilon$ as defined by Eq.(\ref{eigen}).

The above equation has a discrete spectrum when $d=\sqrt{2 K D}/(k_\beta T a) > 1/2$ and
the eigenvalues and the normalized eigen functions of the ground state are then
\equation
\epsilon_0 = \frac{1}{2 \beta} ~ln~ \left ( \frac{\beta K \alpha^2}{2 \pi}~ \right )
             + \frac{a}{\beta} \left ( \frac{D}{2 K} \right )^2 - \frac{a^2}{8 \beta ^2 K},
            \endequation
\equation
\phi_0(y)=\sqrt{a} \frac{(2d)^{d-1/2}}{[\Gamma(2d-1)]^{1/2}} exp(-d e^{-ay})
         exp[-(d-\frac{1}{2})a y ]. \endequation

\end{document}